\documentclass[showpacs,preprintnumbers,amsmath,amssymb]{revtex4}

\usepackage{amsfonts}                                         
\usepackage[dvips]{graphicx}                                  
\usepackage{type1cm}                                          
\usepackage{color}

\newcommand{\eqnref}[1]{Eqn.~\eqref{#1}}                      
\newcommand{\figref}[1]{Fig.~\ref{#1}}                        
\newcommand{\secref}[1]{Sec.~\ref{#1}}                        
\newcommand{\textprog}[1]{#1}                                 
\newcommand{\expecval}[1]{\langle #1\rangle}                  
\newcommand{\Bexpecval}[1]{\left\langle #1\right\rangle}      
\newcommand{\diffd}{\text{d}}                                 
\newcommand{\bracetextsize}{\displaystyle}                    
\renewcommand{\vec}[1]{\mathbf{#1}}                           
\newcommand{\ubrace}[2]{\underbrace{#1}_{\bracetextsize{#2}}} 

\newcommand{\perc}{\,\%}
\newcommand{\unit}[1]{\,\text{#1}}
\newcommand{\punc}[1]{\,#1}

\begin{document}
\title{Diffusion Monte Carlo study of a valley
degenerate electron gas and application to quantum dots}

\author{G.~J. Conduit}
\email{gjc29@cam.ac.uk}
\affiliation{Theory of Condensed Matter, Department of Physics, University of
  Cambridge, Cavendish Laboratory, 19, J.~J. Thomson Avenue, Cambridge, CB3~0HE,
  United Kingdom}
\author{P.~D. Haynes}
\affiliation{Departments of Physics and Materials, Imperial College London,
  Exhibition Road, London, SW7~2AZ, United Kingdom}
\date{\today}

\begin{abstract} A many-flavor electron gas (MFEG) in a semiconductor with a
valley degeneracy ranging between 6 and 24 was analyzed using diffusion Monte
Carlo (DMC) calculations.  The DMC results compare well with an analytic
expression derived by one of us [Phys. Rev. B \textbf{78}, 035111 (2008)] for
the total energy to within $\pm1\perc$ over an order of magnitude range of
density, which increases with valley degeneracy. For
$\text{Bi}_{2}\text{Te}_{3}$ (six-fold valley degeneracy) the applicable charge
carrier densities are between $7\times10^{19}\unit{cm}^{-3}$ and
$2\times10^{20}\unit{cm}^{-3}$. DMC calculations distinguished between an exact
and a useful approximate expression for the 24-fold degenerate MFEG
polarizability for wave numbers $2p_{\text{F}}<q<7p_{\text{F}}$. The analytical
result for the MFEG is generalized to inhomogeneous systems by means of a
gradient correction, the validity range of this approach is obtained. Employed
within a density-functional theory calculation this approximation compares well
with DMC results for a quantum dot. \end{abstract}

\pacs{71.15.Mb,71.10.Ca,02.70.Ss}

\maketitle

\section{Introduction}\label{sec:Introduction}

\begin{figure}
 \includegraphics{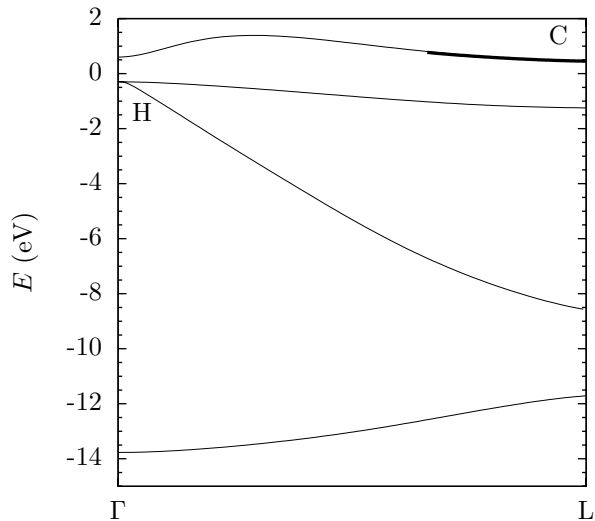}
 \caption{The Ge band-structure in the $[111]$ direction calculated using a
 plane-wave pseudopotential method \cite{05csphprp05}. The Fermi energy is at
 $E=0\unit{eV}$; below are valence bands with the holes centered around H, above
 are conduction bands. The first conduction band valley is highlighted in bold,
 low-lying conduction-band electrons are centered around C.}
 \label{fig:GeBandStructure}
\end{figure}

Good quantum numbers, that describe conserved quantities as a quantum system
evolves, derive their significance from their connection to the powerful
conservation laws of physics.  In addition to the familiar examples of spin and
crystal momentum, under some circumstances electrons in solids can have an
additional quantum number that distinguishes them, which we call the
\emph{flavor}; we denote the total number of flavors by $\nu$. One example of
such a system are semiconductors and semimetals that have degenerate
conduction-band valleys, the flavor denotes the electron's
valley. Examples of multi-valley semiconductors include Ge, which as shown in
\figref{fig:GeBandStructure} has four degenerate valleys (N.B. not eight, as
valleys at the Brillouin zone vertices overlap), Si has six degenerate valleys,
a Ge-Si alloy has ten degenerate valleys, and
$\text{Pb}_{1-x-y}\text{Sn}_{x}\text{Mn}_{y}\text{Te}$ has twelve valleys in the
$\Sigma$ band \cite{90sksg12}. The system has been experimentally realized as an
electron-hole liquid that forms in drops \cite{76abkos08,08c07}. In these systems
the number of flavors (the number of valleys) is well defined and there are
strong Coulomb interactions between particles which motivates the analysis. This
is in contrast to several other systems in which the number of flavors is poorly
defined such as heavy fermions \cite{02zcja03,05k12,06bi02}, charged domain
walls \cite{98eogsz09}, a super-strong magnetic field \cite{76ko07}, and spin
instabilities \cite{86gq04,00mqg03}; or where the number of flavors is well
defined but interactions between particles are weak such as ultracold atoms in
optical lattices \cite{04hh04,04hh09,07crd09}.

The properties of a many-flavor electron gas (MFEG) in a semiconductor were
first studied analytically for the normal phase by \citet{76abkos08}, and for
the superconducting phase by \citet{64c04}. Recently one of us \cite{08c07}
extended the MFEG analysis by finding an energy functional and gradient
expansion, which allowed the study of inhomogeneous systems. However, the
analytical treatment was limited to consider the same contributions to the
energy as in the random phase approximation (these contributions dominate in the
many-flavor limit). To go further requires numerical calculations, the only
example of which for a MFEG to date \cite{94g08} used a self-consistent approach
for the local field correction formulated by \citet{68stls12} (STLS), see also
Ref.~\cite{81st06}. The method was later applied to charge impurities by
\citet{97bat12}. The calculations of Ref.~\cite{94g08} were performed for
$\nu\leq6$, too few flavors to gauge the applicability of the analytic
many-flavor approximation, which is estimated to apply at around six or more
flavors \cite{08c07}.

In this paper we follow the suggestion of \citet{94g08}, and present the results
of what are expected to be more accurate diffusion Monte Carlo (DMC)
\cite{94hlr10,97l07,99nu01,01fmnr01} calculations on the MFEG for $\nu\leq24$,
which should allow us to verify the analytical MFEG approach. We then examine
aspects of the many-flavor approximation that have not yet been studied
computationally: in \secref{sec:CASINOElectronGasDensityResponse} we compare the
analytical density-density response function derived in
\secref{sec:AnalyticalPolarisability} with that predicted using DMC. Once
verified this allows us in \secref{sec:QuantumDotsQMC} to employ a gradient
expansion within density-functional theory (DFT) to find the ground state of a
quantum dot, we compare results with DMC calculations and examine the validity
of the gradient expansion.

We adopt the atomic system of units: that is
$e^{2}=\hbar=m=1/(4\pi\epsilon_{0})=1$. The mass $m=m_{\text{e}}m^{*}$ is
defined to be the electron mass, $m_{\text{e}}$, multiplied by a dimensionless
effective mass $m^{*}$ appropriate for the conduction-band valleys, which when
$m^{*}=1$ will recover standard atomic units. We assume the valleys all have the
same dispersion profile and so the same effective mass, \citet{76abkos08}
outlined a method of calculating a scalar effective mass for anisotropic
valleys. With the above definitions, energy is given in terms of an exciton
$E_{\text{h}}^{*}=E_{\text{h}}m^{*}$, where $E_{\text{h}}$ is the Hartree
energy, and length $a_{0}^{*}=a_{0}/m^{*}$ in terms of the Bohr radius
$a_{0}$. To denote density we use both the number density of conduction-band
electrons $n$ and the Wigner-Seitz radius $r_{\text{s}}$.

Before presenting the numerical results, to orient the discussion, we describe
the basic physics of the MFEG and review the analytical results of
Ref.~\cite{08c07} that will be computationally verified in this paper.

\subsection{Introduction to a MFEG}\label{sec:AnalyticalResults}
 
In a low temperature MFEG, the number of flavors $\nu$, number density of
conduction-band electrons $n$, and Fermi momentum $p_{\text{F}}$ are related
through
\begin{equation}
 \label{flavoursdensityofstates}
 n=\frac{\nu p_{\text{F}}^{3}}{3\pi^{2}}\punc{.}
\end{equation}
At fixed electron density, the Fermi momentum reduces with increasing number of
flavors as $p_{\text{F}}\propto\nu^{-1/3}$, so each Fermi surface encloses fewer
states. The semiconductor hole band-structure often has a single valence-band
minimum at the $\Gamma$ point, such as in Ge, see \figref{fig:GeBandStructure},
hence we assume the holes are heavy and are uniformly distributed, providing a
jellium background.

For a constant number density of particles, the density of states at the Fermi
surface, $g$, rises with increasing number of flavors as
$g\propto\nu\sqrt{E_{\text{F}}}\propto\nu^{2/3}$. Therefore, the screening
length estimated with the Thomas-Fermi approximation \cite{04k11} is
$\kappa^{-1}=(4\pi e^{2}g)^{-1/2}\propto\nu^{-1/3}$, and the ratio of the
screening to Fermi momentum length-scale varies with number of flavors as
$p_{\text{F}}/\kappa\propto\nu^{-2/3}$. In the many-flavor limit $\nu\gg1$, the
screening length is much smaller than the inverse Fermi momentum,
$\kappa^{-1}\ll p_{\text{F}}^{-1}$, and so the dominant electron-electron
interactions have characteristic wave vectors which obey $q\gg p_{\text{F}}$.
This is in direct contrast to the random phase approximation (RPA) where
$p_{\text{F}}\gg\kappa$, although in both the many-flavor and the RPA, the same
Green function contributions with empty electron loops dominate diagrammatically
\cite{08c07,76abkos08}. These diagrams contain the greatest number of different
flavors of electrons, and as $\nu\gg1$ therefore have the largest matrix
element. Since $q\gg p_{\text{F}}$, the typical length-scales of the MFEG are
short, this indicates that a local density approximation (LDA) could be
applied. This motivation is in addition to the usual reasons for the success of
the LDA in DFT \cite{92ptaaj10}, namely that the LDA exchange-correlation hole
need only provide a good approximation for the spherical average of the
exchange-correlation hole and obey the sum rule \cite{89jg06}.

\subsection{Polarizability}\label{sec:AnalyticalPolarisability}

In the many-flavor limit the exact result for the polarizability of a MFEG at
wave vector $\vec{q}$, and Matsubara frequency $\omega$ is 
\cite{76abkos08,78br07,08c07}
\begin{widetext}
 \begin{eqnarray}
  \label{refeqnpolarisability}
  \Pi_{0}(\omega,q)&=&\frac{\nu}{2\pi^{2}}\Biggl[\frac{\omega}{q}\left({\text{tan}}^{-1}\left(\frac{
q/2+p_{\text{F}}}{\omega/q}\right)-{\text{tan}}^{-1}\left(\frac{
q/2-p_{\text{F}}}{\omega/q}\right)\right)\nonumber\\
&-&\frac{\left(\omega/q\right)^{2}+p_{\text{F}}^2-q^{2}/4}{2q}{\ln}\left(\frac{\left(\omega/q\right)^{2}+\left(q/2+p_{\text{F}}\right)^{2}}{\left(\omega/q\right)^{2}+\left(q/2-p_{\text{F}}\right)^{2}}\right)-p_{\text{F}}\Biggr]\punc{,}
 \end{eqnarray}
\end{widetext}
which in the many-flavor limit is approximately
\begin{equation}
 \Pi_{0}(\omega,q)=-\frac{n}{(\omega/q)^2+q^{2}/4}+\mathcal{O}(\nu^{-2/3})\punc{.}
 \label{refeqnpolarisabilityexpansion}
\end{equation}
This quantity governs the density-density response of the MFEG so is important
to verify. Since \eqnref{refeqnpolarisabilityexpansion} has a simple form it
can be used to calculate further properties of the MFEG \cite{08c07}, such as
homogeneous energy in \secref{sec:AnalyticalUniformEnergy} and the gradient
expansion in \secref{sec:AnalyticalGradient}, which further motivates its
numerical verification.

\subsection{Homogeneous energy}\label{sec:AnalyticalUniformEnergy}

Starting from the approximate expression for polarizability,
\eqnref{refeqnpolarisabilityexpansion}, it can be shown that the total energy of
a MFEG, including all the exchange and correlation contributions is \cite{08c07}
\begin{equation}
 E=\frac{3}{10}\left(\frac{3\pi^{2}}{\nu}\right)^{2/3}n^{5/3}-\ubrace{A_{\text{3D}}n^{5/4}}{E_{\text{int}}}\punc{,}
 \label{eqn:CASINOHomogeneousFreeElectronGasTotalEnergy}
\end{equation}
where $A_{3\text{D}}=\Gamma(-5/4)\Gamma(3/4)/(2\pi^{5/4})(E_{\text{h}}^{*}
a_{0}^{*3/4})\approx0.574447(E_{\text{h}}^{*} a_{0}^{*3/4})$ and
$E_{\text{int}}$ denotes the interacting energy (which would be zero if
electron-electron interactions were ignored).

In Ref.~\cite{08c07} it was suggested that this relation for the total energy
applies over a density range, at $99\perc$ accuracy, $0.03\nu\ll
na_{0}^{*3}\ll(0.074\nu)^{4}$, which widens with number of flavors as $\nu^{4}$
(see also Ref.~\cite{76abkos08}). Considering the number of flavors where the
range of validity vanishes indicates that the many-flavor limit will apply if
there are ten or more flavors. An alternative estimate for the density range is
found in \secref{sec:GroundStateEnergyVariation} by comparing the analytical
result with DMC calculations.

\subsection{Gradient correction}\label{sec:AnalyticalGradient}

The applicability of the LDA in a MFEG motivates the search for a gradient
expansion to the energy \eqnref{eqn:CASINOHomogeneousFreeElectronGasTotalEnergy}
as a way to analyze inhomogeneous systems such as electron-hole drops and
quantum dots. The typical momentum transfer in the MFEG is $q\sim4(\hbar
a_{0}^{*-1/4})n^{1/4}$, which defines the shortest length-scale over which a LDA
can be made, therefore, the maximum permissible gradient in electron density is
$|\nabla n|_{\text{max}}\sim qn\sim4(\hbar a_{0}^{*-1/4})n^{5/4}$. A gradient
expansion will break down for phenomena with short length-scales, for example
mass enhancement \cite{93m08}. If electron density is smoothly varying then
starting from \eqnref{refeqnpolarisabilityexpansion}, the gradient correction to
the energy for a MFEG is \cite{08c07}
\begin{equation}
 E=E_{0}+\frac{1}{8}\frac{(\nabla n)^{2}}{n}\punc{,}
 \label{eqn:GradientExpansion}
\end{equation}
where $E_{0}$ is the energy of a homogeneous MFEG with density $n$,
see \eqnref{eqn:CASINOHomogeneousFreeElectronGasTotalEnergy}. As discussed in
\secref{sec:AnalyticalResults}, this gradient expansion would be useful for DFT
calculations and so its computational verification is important.

\section{Computational method}

In this section we briefly describe the two computational methods that we used,
variational Monte Carlo (VMC) and Diffusion Monte Carlo (DMC)
\cite{01fmnr01}. These are quantum Monte Carlo (QMC) methods, chosen since DMC
gives the exact ground state energy subject to the fixed node approximation, and
both are expected to give more accurate results than the STLS approach used by
\citet{94g08}.

The VMC method uses a normalizable and differentiable trial wave function
$\Psi_{\text{T}}$, of the form discussed below. The Metropolis algorithm
\cite{53mrtt06} is used to sample the wave function probability density
$|\Psi_{\text{T}}|^{2}$ using a random walk, and make an estimate of the local
energy
$E_{\text{L}}(\vec{r})=\Psi_{\text{T}}(\vec{r})^{-1}\hat{H}\Psi_{\text{T}}(\vec{r})$.
In order to obtain the ground state one could minimize the spatial average of
the local energy with respect to the free parameters in the trial wave
function. However, it is computationally more stable to minimize the variance in
the estimates of the local energy. As VMC obeys the variational principle by
construction, it yields an upper bound to the true ground state energy.

The more accurate DMC algorithm is a stochastic method that begins with a trial
or guiding wave function, in this case the optimized VMC trial wave
function. The DMC method is based on imaginary time evolution, which when using the
operator $\text{e}^{-t(\hat{H}-E_{\text{T}})}$ projects out the ground state
wave function from the trial wave function, and yields an estimate of the ground state
energy, $E_{\text{T}}$. The nodal surface on which the wave function is zero (and
across which it changes sign) is fixed \cite{82rcal12,99nu01} to be that of the
trial wave function, this ensures that the fermionic exchange symmetry is
maintained. The DMC algorithm produces the exact ground state energy subject to the
fixed node approximation, and is also variational so gives an accurate upper
bound to the true ground state energy once the population control bias and finite
time-step bias are eliminated. The algorithm used closely follows that
described in Ref.~\cite{93unr08}.

In our QMC calculations we use a Slater-Jastrow \cite{55j06,01fmnr01,04dtn12}
trial wave function. The Slater part of the wave function is a product of
determinants, each one corresponding to a different electron spin or
flavor. Each determinant is over the spatial orbitals of electrons occupying the
lowest energy levels. The determinant changes sign when rows or columns are
swapped, this ensures that the wave function is antisymmetric under exchange of
electrons with the same flavor and spin. The Slater wave function itself is not
the ground state of an interacting electron gas, so to improve the wave
function, variational degrees of freedom that account for two-body correlations
are included within a Jastrow factor. The Jastrow factor is symmetric under
particle exchange so does not alter the particle exchange symmetry of the wave
function.  Furthermore, the Jastrow factor is always positive so does not alter
the wave function nodal surface. The Jastrow factor contains a two-body
polynomial term $u(r_{ij})=F(r_{ij})\sum_{l=2}^{6}\alpha_{l}r_{ij}^{l}$, a power
series form \cite{04dtn12} in electron separation $r_{ij}$ with optimizable
parameters, $\alpha_{l}$.  The term $F(r_{ij})$ ensures that the Kato cusp
conditions are satisfied \cite{66pb02}. To ensure that electron-electron
correlations do not extend beyond the simulation cell, the term is cutoff at the
Wigner-Seitz radius. To treat longer-ranged correlations, the Jastrow factor
includes a two-body plane-wave expansion,
$p(\vec{r}_{ij})=\sum_{A,\vec{G}_{A}}a_{A}\cos(\vec{G}_{A}\cdot\vec{r}_{ij})$.
Those reciprocal lattice vectors, $\{\vec{G}_{A}\}$, that are related by the
point group symmetry (denoted by $A$) of the Bravais lattice share the same
optimizable parameters, $a_{A}$. To ensure accuracy we checked the stability of
the VMC results when the expansion order of the $u$ and $p$ terms was increased.
At all densities the Jastrow factor optimized cutoff lengths took the maximum
allowed value (the Wigner-Seitz radius).

The DMC calculations were performed with $57$ different reciprocal lattice
vectors and, following \citet{94ob07}, \citet{78c10}, and \citet{80ca08},
further VMC calculations were performed at other system sizes (27, 33, 57, and
81 reciprocal lattice vectors) to derive the parameters to extrapolate the DMC
energy to infinite system size. Additionally, all the DMC results were
extrapolated to have zero time-step between successive steps in the electron
random walk. In DMC simulations the acceptance probability of a proposed step in
the random walk exceeded $99\perc$. We used $300$ DMC configurations, comparable
to the 200-300 used by \citet{94ob07}, and checked for population control bias
by ensuring that ground state energy estimates did not vary with a changing
number of configurations. All the QMC calculations were performed using the
\textprog{CASINO} computer program \cite{06ntdl07}.

\section{Homogeneous MFEG}\label{sec:CASINOHomogeneousFreeElectronGas}

We start with the simplest possible system to analyze numerically, the
homogeneous MFEG, this provides not only a suitable system to validate both
theory (\secref{sec:AnalyticalUniformEnergy}) and the QMC many-flavor
calculations, but should also confirm the range of densities over which the
many-flavor approximation applies. The 3D homogeneous electron gas ($\nu=1$) has
been studied before using QMC \cite{78c10,80ca08,94ob07} and these studies
provide a useful guide to the method we should follow.

\begin{figure}
  \includegraphics{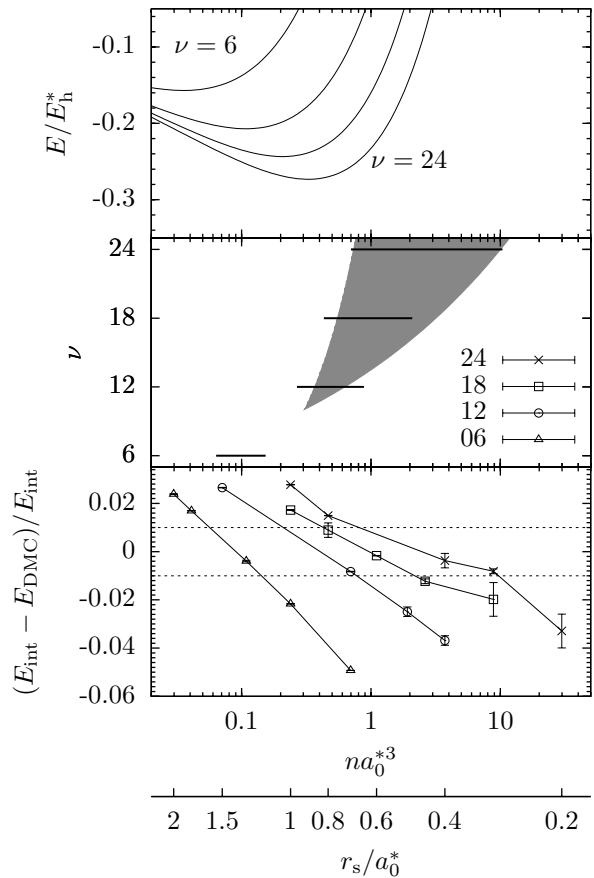}
  \caption{The lower panel shows the fractional difference of DMC interaction
  energy $E_{\text{DMC}}$ from the model $E_{\text{int}}$ with MFEG density $n$
  (and Wigner-Seitz radius $r_{\text{s}}$) for different numbers of flavors, the
  dotted lines show $\pm1\%$ disagreement. The central panel bars
  highlight the numerical region of applicability, the gray shaded area
  represents the analytically predicted region of $\pm1\%$ applicability. The
  upper panel shows the total energy for 6, 12, 18 and 24 flavor electron gases.
  }
 \label{fig:CASINOhomogeneousgas}
\end{figure}

To calculate the interaction energy $E_{\text{int}}$ we subtracted the
theoretical Thomas-Fermi kinetic energy from the DMC ground state energy (see
\eqnref{eqn:CASINOHomogeneousFreeElectronGasTotalEnergy}). At each of 6, 12, 18,
and 24 flavors we performed five DMC calculations and interpolated to find where
theory and DMC results agree to within $\pm1\%$. Results in
\figref{fig:CASINOhomogeneousgas} show that for $\nu>6$ the theory applies over
at least an order of magnitude in density to an accuracy of $\pm1\perc$ -- the
theory can be applied at fewer flavors than expected. For fewer than $\sim12$
flavors the valid logarithmic range of the theory increases with $\nu$, the $18$
and $24$ flavor results show a dramatic increase in the range of validity,
especially on the high density side. In the limit of many flavors $(\nu>12)$ the
expected $99\perc$ range of validity $0.03\nu\ll na_{0}^{*3}\ll(0.074\nu)^{4}$
is approximately consistent with the computationally predicted $\pm1\perc$
region, therefore the minimum number of flavors required for all aspects of
the many-flavor theory to be valid is approximately ten.

For Si with $m^{*}=1.08$ the many-flavor limit applies to an accuracy of
$\pm1\perc$ for a charge carrier concentration between
$4\times10^{23}\unit{cm}^{-3}$ and $1\times10^{24}\unit{cm}^{-3}$, this is
greater than the typical maximum carrier density
$\sim1\times10^{21}\unit{cm}^{-3}$ and so in Si the formalism is not applicable.
In systems with a low effective mass, for example the $\nu=6$ material
$\text{Bi}_{2}\text{Te}_{3}$ used in thermoelectric cooling, which has
$m^{*}=0.06$ \cite{98hhosk06,06la01,08nkopkrvkjhl03}, the required charge
carrier concentration is between $7\times10^{19}\unit{cm}^{-3}$ and
$2\times10^{20}\unit{cm}^{-3}$, which compares favorably with the typical
maximum carrier density $\sim1\times10^{21}\unit{cm}^{-3}$ and so the
many-flavor limit formalism could be applied to low effective mass
materials.

The STLS results of \citet{94g08} at $\nu=1$ were $\sim3.4\%$ less negative than
the DMC results of \citet{94ob07}, and at $\nu=6$ were $\sim3.1\%$ less negative
than our DMC results. This represents a significant difference between our and
the STLS results when looking for the $1\%$ range of validity, highlighting the
need for the more accurate DMC calculations. The range of validity at
$\pm1\perc$ up to at least 24 flavors is to the high density side of the minimum
in the total energy seen in \figref{fig:CASINOhomogeneousgas}, but the minimum
$n_{\text{min}}\propto\nu^{8/5}$ lies within the region of validity for higher
$\nu$. $n_{\text{min}}$ is the density expected to be seen in physical systems
such as electron-hole drops, the good agreement of the theory with DMC results
at this density indicates that the theory could be usefully applied to
investigate the properties of physical systems, see for example
Ref.\cite{08c07}.

\section{Static density-density response}\label{sec:CASINOElectronGasDensityResponse}

Having verified the homogeneous system behavior we may now proceed and
computationally examine inhomogeneous behavior through the static
density-density (linear) response function
\eqnref{refeqnpolarisabilityexpansion}. The polarizability is an important
quantity used \cite{08c07} to develop both homogeneous theory and the gradient
correction, the density-density response function itself also governs the
electrical response properties, for example polarization, screening, and
behavior in an external potential; it is therefore useful to verify this
response before applying the theory to model systems. We examine
$1/\epsilon(\vec{q})$, the quantity probed experimentally \cite{99np10}.

DMC has previously been used to find the static density-density response of
single-flavor systems: \citet{92sba11} applied the method to charged bosons, the
density-density response of the electron gas was calculated by \citet{92mcs09}
(in two dimensions), and \citet{94bsa11} and \citet{95mcs07} (three dimensions).
However, density-density response has not been studied numerically in a
many-flavor system. Here we employ two methods to find the density-density
response function. The more accurate and computationally efficient method of
calculating the response is to examine the ground state energy, calculated using
DMC.  A VMC energy based estimate and an estimate using the induced electron
density are used to check the accuracy of the trial wave function.

Before the results are described in
\secref{sec:StaticDielectricResponseResults}, we outline the theory behind the
two methods used to estimate the response, firstly in
\secref{sec:GroundStateEnergyVariation} by using the ground state energy
variation, and secondly in \secref{sec:InducedChargeDensityMeasurement} through
the magnitude of the periodic density modulation.

\subsection{Ground state energy variation}\label{sec:GroundStateEnergyVariation}

To calculate the density-density response we use a weak probe so that the
density response is solely due to the properties of the homogeneous system. We
apply a static ($\omega=0$) monochromatic perturbative external potential
$U(\vec{r})=U_{\vec{q}}\cos(\vec{q}\cdot\vec{r})$ to the homogeneous MFEG,
corresponding to the background charge having an additional sinusoidal variation
$n_{\text{ext}}(\vec{r})=n_{\vec{q}}\cos(\vec{q}\cdot\vec{r})$. The external
potential and external charge are linked \cite{92sba11} through Poisson's
equation by
\begin{equation}
 \label{eqn:ExternalChargeDensity}
 n_{\text{ext}}(\vec{k})=\frac{U_{\vec{q}}q^{2}}{8\pi}(\delta_{\vec{k},\vec{q}}+\delta_{\vec{k},\vec{-q}})\punc{.}
\end{equation}
We assume that different Fourier components are independent, the density
response to an external potential with wave vector $\vec{q}$ and frequency
$\omega$ is only at that wave vector and frequency so the induced charge is
$n_{\text{ind}}(\vec{k})=\left(\Bexpecval{\hat{n}_{\vec{k}}}_{U_{\vec{q}}}-\Bexpecval{\hat{n}_{\vec{k}}}_{0}\right)(\delta_{\vec{k},\vec{q}}+\delta_{\vec{k},\vec{-q}})$.
Here $\expecval{\hat{n}_{\vec{k}}}_{U_{\vec{q}}}$ is the expectation value of
the charge density Fourier component at wave vector $\vec{k}$ with an applied
external potential $U_{\vec{q}}$, and $\expecval{\hat{n}_{\vec{k}}}_{0}$ is the
same but in the homogeneous case with no external potential. Linear response
theory gives the static density-density response function as the ratio of the
induced charge density and the perturbing external charge density so
\begin{equation}
 \label{eqn:StaticDielectricFunctionDefinition}
 \frac{1}{\epsilon(\vec{q})}=1+\frac{8\pi}{U_{\vec{q}}q^{2}}\left(\Bexpecval{\hat{n}_{\vec{q}}}_{U_{\vec{q}}}-\Bexpecval{\hat{n}_{\vec{q}}}_{0}\right)\punc{.}
\end{equation}
If the external potential is small relative to other typical energies the
density response is determined solely by the properties of the homogeneous
MFEG. We can expand in small $U_{\vec{q}}$ so that
\begin{eqnarray}
 \Bexpecval{\hat{n}_{\vec{k}}}_{U_{\vec{q}}}-\Bexpecval{\hat{n}_{\vec{k}}}_{0}&\approx&\left.U_{\vec{q}}\frac{\diffd\Bexpecval{\hat{n}_{\vec{k}}}}{\diffd U_{\vec{q}}}\right|_{U_{\vec{q}}=0}=\left.U_{\vec{q}}\frac{\diffd ^{2}E}{\diffd U_{\vec{q}}^{2}}\right|_{U_{\vec{q}}=0}\punc{,}
\end{eqnarray}
where the induced charge density is calculated by considering the dependence of
the ground state energy $E$ on the magnitude of the external field. Substituting
this into \eqnref{eqn:StaticDielectricFunctionDefinition} gives an expression
for the density-density response
\begin{equation}
 \frac{1}{\epsilon(\vec{q})}=1+\frac{8\pi}{q^{2}}\left.\frac{\diffd ^{2}E}{\diffd U_{\vec{q}}^{2}}\right|_{U_{\vec{q}}=0}\punc{.}
\end{equation}

To recover the density-density response function at a particular wave vector,
several QMC calculations were performed at that wave vector for different
amplitudes of the external field. A polynomial fit was made to the ground state
energy so as to extract the second derivative. To investigate the lowest order
polarizability the applied external field should be as small as possible yet
still give statistically significant results, to ensure this we checked that the
ground state energy showed only quadratic behavior with applied field amplitude.
A further convenient way to check the perturbing field is sufficiently small is
to ensure the electric field of the external potential is less than the typical
electric field strength between two neighboring electrons, $e/r_{\text{s}}^{2}$.

\subsection{Induced charge density measurement}\label{sec:InducedChargeDensityMeasurement}

As the external potential is perturbative we use the same plane-wave basis set
as employed for the calculations on the homogeneous MFEG described in
\secref{sec:CASINOHomogeneousFreeElectronGas}. To account for the modulating
density, following \citet{92mcs09}, \citet{94bsa11}, and \citet{95mcs07} we
introduce a new $q$ term into the Jastrow factor of the form
\begin{equation}
 \label{eqn:PolarisabilityJastrowqTerm}
 q(\vec{r}_{i})=b\cos(\vec{q}\cdot\vec{r}_{i})\punc{,}
\end{equation}
where $b$ is an optimizable parameter, $\vec{r}_{i}$ the position of the $i$th
electron, and the wave vector $\vec{q}$ corresponds to that of the perturbative
external potential. As $b$ is small, the charge density induced by the
perturbative external potential is
$n_{\text{ind}}\approx2b\cos(\vec{q}\cdot\vec{r}_{i})$. From
\eqnref{eqn:ExternalChargeDensity} and
\eqnref{eqn:StaticDielectricFunctionDefinition} it follows that
\begin{equation}
 \label{eqn:InversePermittivityFromJastrowCoefficient}
 \frac{1}{\epsilon(\vec{q})}=1+\frac{8\pi b}{q^{2}U_{\mathbf{q}}}\punc{.}
\end{equation}
The optimized value of $b$ was found by variance minimization during a VMC
calculation. The relationship then allows us to derive an estimate for the
density-density response function for each separate $U_{\vec{q}}$, typically
four values were averaged to give a final estimate for the density-density
response.

\subsection{Results}\label{sec:StaticDielectricResponseResults}

We chose to find the polarizability for a MFEG with $\nu=24$ and
$r_{\text{s}}=0.6a_{0}^{*}$. This lies at the lower bound of the range of
validity near to the minimum in the energy (see
\figref{fig:CASINOhomogeneousgas}) at a density expected to be seen in physical
systems. This density was also chosen since it had most of the polarizability
curve $0.25<1/\epsilon\le1$ in the region of applicability
$q\ge2p_{\text{F}}$. Boundary conditions mean that the external potential must
be periodic over the simulation cell, therefore the external potential wave
vector $\vec{q}$ must be a reciprocal lattice vector. We checked that if the
Jastrow factor $q$ term wave vector was changed so that it was incommensurate
with the external potential then following optimization $b=0$ within statistical
errors; this verified the linear response assumption that Fourier components are
independent.

\begin{figure}
 \includegraphics{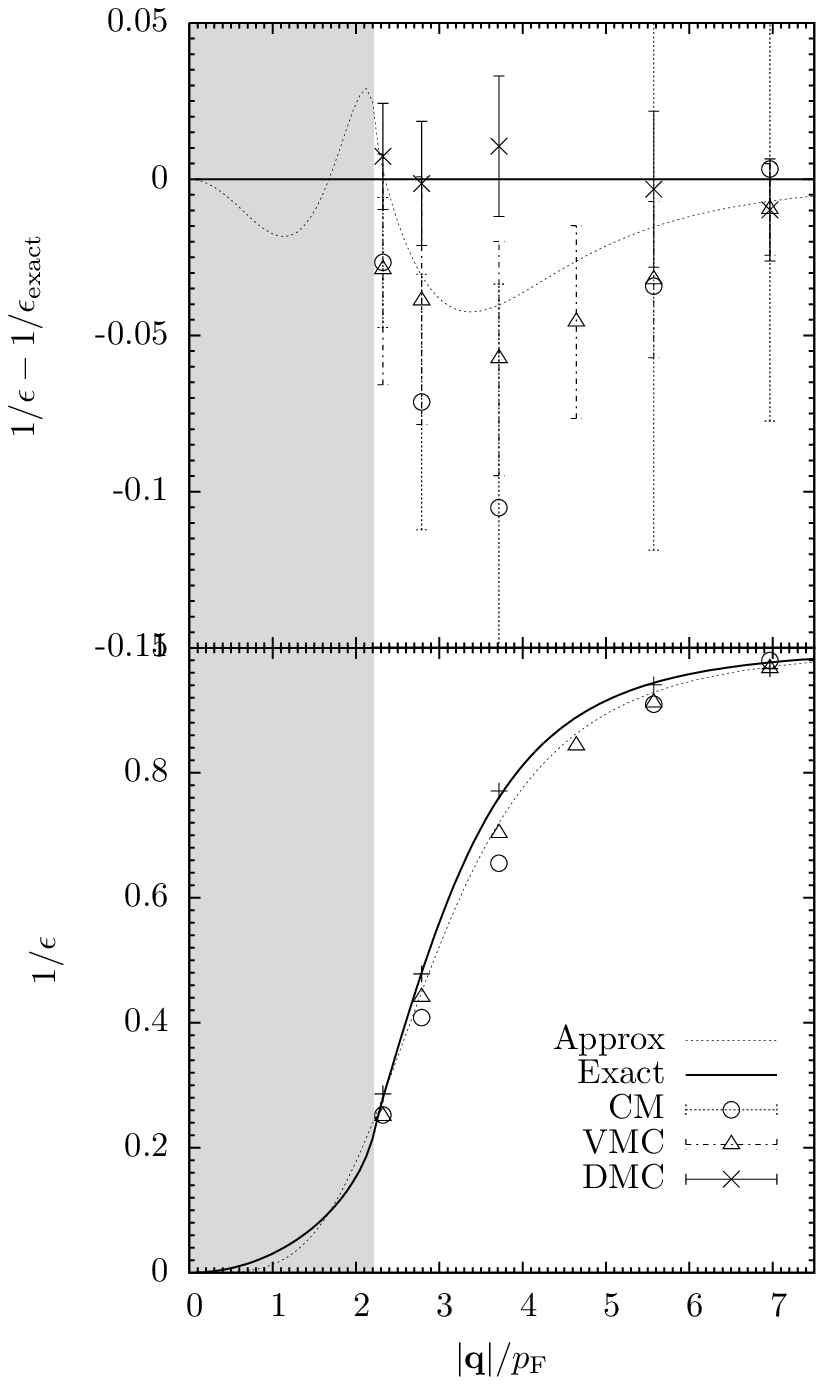}
 \caption{The density-density response $1/\epsilon$ versus the wave vector
 $|\vec{q}|$ of a MFEG with $\nu=24$ and $r_{\text{s}}=0.6a_{0}^{*}$. The
 solid curve shows the exact result $1/\epsilon_{\text{exact}}$
 (\eqnref{refeqnpolarisability}), the dotted curve the
 \eqnref{refeqnpolarisabilityexpansion} approximation. The shaded grey region
 $|\vec{q}|/2<p_{\text{F}}$ is where the many-flavor limit breaks down. The
 points show the values for the permittivity calculated from QMC results, the
 circle is from modulated charge measurements (CM), the triangle from VMC energy
 and the cross from DMC energy. The lower panel plots the actual response, the
 upper panel shows the deviation of response from the exact theoretical result
 \eqnref{refeqnpolarisability} with standard error bars.}
 \label{fig:CASINOpolarisability}
\end{figure}

The results of the calculation are shown in
\figref{fig:CASINOpolarisability}. The DMC results obtained by considering the
variation in ground state energy (see \secref{sec:GroundStateEnergyVariation})
better fits the exact than approximate expression for the polarizability, and
though error bars are large can distinguish between the two within one standard
deviation. This shows that QMC results can exceed the accuracy of the
approximation made in \eqnref{refeqnpolarisabilityexpansion}, though that
estimate remains useful. The positive agreement verifies the theory and confirms
the accuracy of the \textprog{CASINO} simulations.

The ground state energies calculated by VMC were used in the same way as the DMC
results to find the density response and provide a reasonable fit, though here
error bars are large so comparison is difficult. Following the prescription in
\secref{sec:InducedChargeDensityMeasurement} we also derived values for the
density-density response function using the charge density modulation at the
wavelength of the perturbing potential, $U_{\vec{q}}$. These values agreed
within statistics though carried a larger uncertainty than those derived using
the ground state energy. Both of these alternative methods appear to
underestimate the density-density response. These results are consistent, a
smaller charge density response gives a smaller coefficient in the Jastrow
factor $q$ term and a smaller reduction in ground state energy. Nevertheless,
the reasonable agreement of both VMC estimates and to the DMC results indicates
that the trial wave function had an adequate nodal surface.

\section{Gradient correction}\label{sec:QuantumDotsQMC}

It was important to verify the density-density response as it is a key component
to the many-flavor formalism and could be applied to other many-flavor systems
where density is expected to be inhomogeneous, for example junctions and the
response to defects and impurities. Now that it has been verified, we may
proceed to consider a quantity derived from it: the gradient expansion,
\eqnref{eqn:GradientExpansion}, which is also useful for analyzing systems with
inhomogeneous density. Once we have investigated the validity of such an
expansion we can apply the formalism to quantum dots, chosen since they have a
large controllable variation in electron density so should provide a good test
of the gradient expansion. Quantum dots are commonly made in many-flavor
semiconductor materials so can be modeled using a many-flavor formalism, and are
a system in which there is current research interest.

Quantum dots \cite{93bk03,02rm10} have not previously been studied in the
many-flavor limit though there have been several previous computational studies
of a single-flavor electron gas confined in a quantum dot. Previous QMC
simulations of quantum dots include \citet{91pk05}, \citet{99hsnh02} performed
VMC calculations for parabolically confined electrons in circular
dots. \citet{96b08} performed fixed-phase DMC simulations. Path-integral QMC
calculations have also been performed \cite{92lm06,98mew11,99ehmg04}, these
showed poor agreement with results from exact
diagonalization~\cite{97m03}. \citet{03bpwggkrp09, 02wghpg11, 03pwrg10} all
compared the optical band-gap between DMC calculations and results from other
methods. For circular quantum dots \citet{00pul09} found the ground-state using
both DMC, a local spin density approximation method, and Hartree-Fock, they then
directly compared the ground-state energy, correlation energy, and spin density
profiles. \citet{06gguub05} also used DMC to investigate circular quantum
dots. Quantum dots have successfully been investigated using DFT
\cite{94fv11,97kmr08,99hw02,00pul09}, \citet{00pul09} found the local spin
density approximation method predicted ground-state energies that were typically
$2\perc$ greater than DMC energies, \citet{94fv11} obtained a $3\perc$ agreement
between current-density-functional theory and exact diagonalization results.

\subsection{Method}

Before describing the study of quantum dots using a many-flavor functional in
detail we first outline the general strategy of the numerical calculations.
Firstly, a DFT calculation using the many-flavor functional (including the
gradient approximation) was performed using a plane-wave basis set. This
produced an estimate of the ground-state energy and density according to the
many-flavor theory. It also provided a trial wave function that was converted to
a B-spline basis set and, with Jastrow factor, was optimized in a VMC calculation,
in preparation for a DMC calculation. Finally, the DMC calculation gave a second
estimate of the ground state energy and density, exact only for the fixed node
approximation. This estimate was compared with the DFT calculation, and
also gave an insight into the accuracy of the many-flavor theory.

Here we carried out simulations on a quantum dot with a harmonic external
potential of the form $V=kr^{2}/2$, where $r$ is the distance to the center of
the quantum dot containing a MFEG with 12 flavors. This potential was chosen as
it is simple, continuous, realistic \cite{96tahhk10,98sat12}, and has been used
in previous computational studies
\cite{94fv11,96a02,97kmr08,97kodeaht12,99hsnh02,99hw02,00pul09,03r03}. Filled
shells in this potential correspond to $1, 4, 10, 20, 35,\dots$ orbitals (whose
degeneracy may be reduced by electron-electron interactions). In DFT we used a
supercell containing a single dot to model the aperiodic system with periodic
boundary conditions, in DMC non-periodic calculations with just a single quantum
dot were performed. The cubic cell was large enough that the trial wave
functions had reduced by at least a factor of $10^{-4}$ at its boundary.

Trial wave functions were generated using the DFT program \textprog{3Ddotdft},
an extended version of \textprog{DOTDFT} \cite{05h01}. This used the many-flavor
functional with gradient approximation so had energy density
\begin{equation}
 \label{eqn:QntmDotDFTFunctional}
 \varepsilon(n(\vec{r}))=-A_{3\text{D}}n^{5/4}+\xi\frac{|\nabla n|^{2}}{8n}\punc{.}
\end{equation}
A new parameter $\xi$ was introduced that multiplies the gradient term, which
allowed us to adjust its size; $\xi=1$ gives the correct analytical expression,
and $\xi=0$ the functional without a gradient expansion.

The VMC simulations, run in \textprog{CASINO}, used a B-spline basis set
\cite{97hgg05,04ag10} because a localized basis set offers
significant performance advantages over plane-waves. The wave function was
optimized in VMC with a Jastrow factor containing the two-body polynomial $u$
term and two-body plane-wave term $p$ with the same form as used in
\secref{sec:CASINOHomogeneousFreeElectronGas} and
\secref{sec:CASINOElectronGasDensityResponse}, and a one-body electron-potential
term $\chi(r_{i})=F(r_{i})\sum_{m=2}^{6}\beta_{m}r_{i}^{m}$ with $F$ determining
behavior at the cutoff length, $r_{i}$ the distance of the $i$th electron from
the center of the potential, and the $\beta_{m}$ being optimizable parameters;
we also note that the $\chi$ term has no central cusp.

The many-flavor functional incorrectly adds in the \emph{self-interaction}
energy of each electron to its own Coulomb potential.  One way to correct for
this is to add an additional term to the density functional
\cite{81pz05,01dg01}. However, as the number of flavors is increased the ratio
of the correct interaction ($\propto\nu^{2}$) to incorrect self-interaction
($\propto\nu$) increases as $\sim2\nu-1$ so in the many-flavor limit the
self-interacting energy error may be neglected. To ensure the B-spline grid was
sufficiently fine, we compared the trial wave function kinetic and external
potential energy before and after conversion the B-spline basis set. We also
checked the choice of DMC time-step was sufficiently small, the number of
configurations was suitably large, and the simulation cell size was adequately
large. On changing these variables the variation in the ground state energy was
$\Delta E\approx0.02E_{\text{h}}^{*}$, sufficiently small to allow us to compare
the ground state energy as the potential strength and gradient expansion
coefficient were varied.

\subsection{Results}

We analyzed a quantum dot containing a MFEG of 12 flavors and 4 bands (shells),
containing a total of 96 electrons. This was chosen since it had a full shell so
is expected to have a zero spin ground-state \cite{02rm10} that can be analyzed
with the many-flavor functional, was computationally feasible, and contained
enough electrons to be in the LDA regime, where the many-flavor functional is
expected to apply.

Two different investigations were carried out to probe effects of changing the
density gradient, firstly strength of the dot confining potential $k$ was
changed, and secondly the gradient expansion coefficient $\xi$ was varied.

\subsubsection{Varying the external potential strength $k$}

\begin{figure}
 \includegraphics{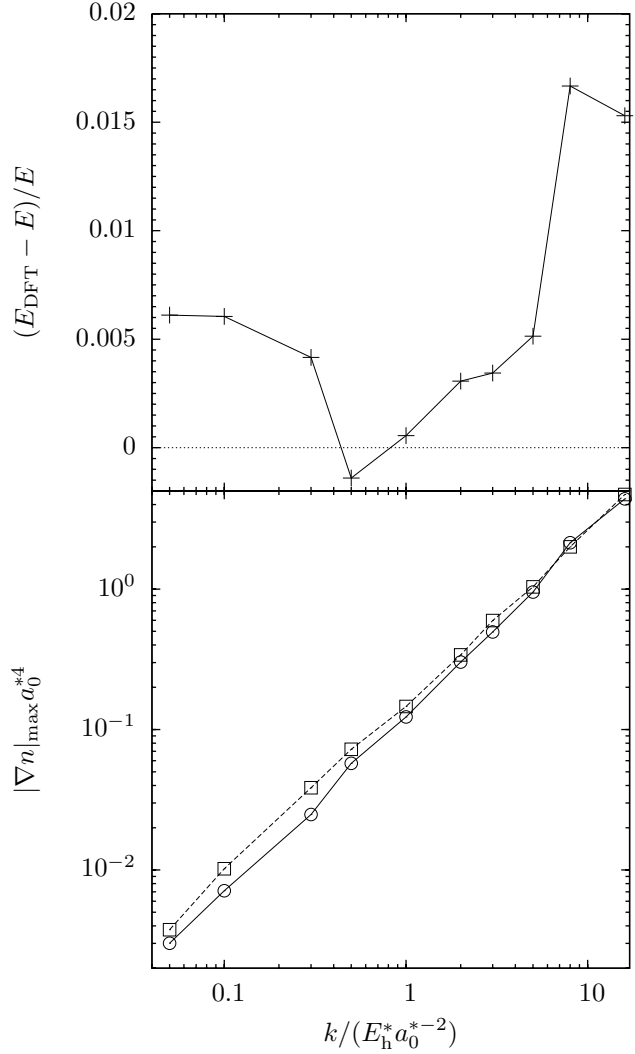}
 \caption{In the upper plot the crosses and solid line show the
 difference between DFT $(E_{\text{DFT}})$ and DMC $(E)$ energies with varying
 external potential strength $k$, if agreement were exact, points would lie
 along the horizontal dotted line. In the lower plot the circles and solid line
 show the maximum density gradient of the dots with varying $k$, and the squares
 and dashed line the gradient at which the theory breaks down.}
 \label{fig:QntmDotVaryK}
\end{figure}

\begin{figure}
 \includegraphics{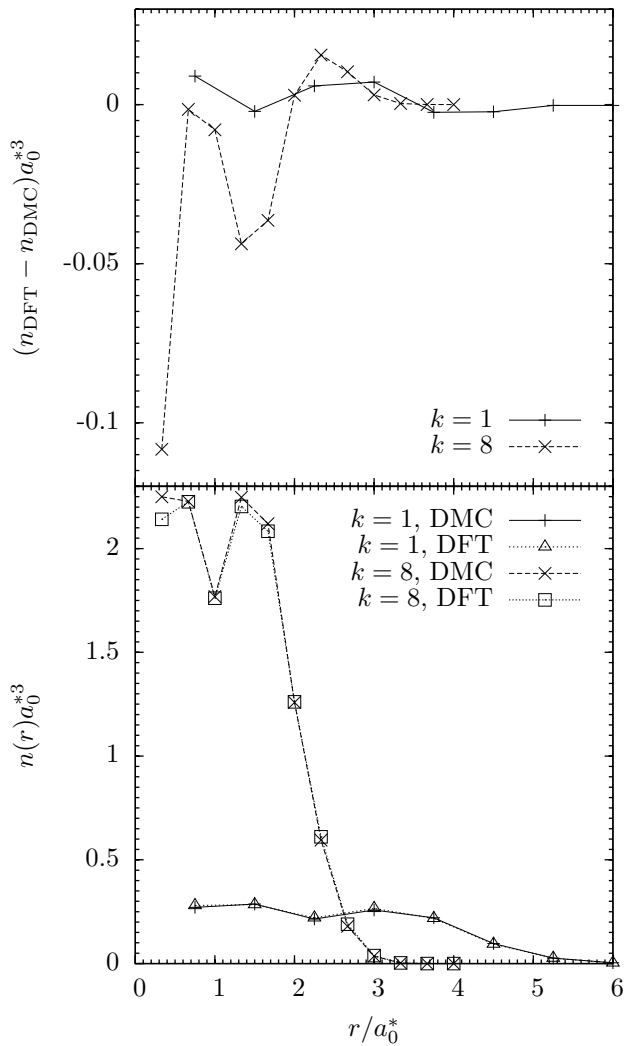}
 \caption{The lower panel shows the density profile of quantum dots estimated
 using both DFT and DMC at external potential strengths of $k=1$ and $k=8$. The
 difference between the DFT and DMC results at $k=1$ and $k=8$ is shown in the
 upper panel. The DMC statistical error is less than the size of the points.}
 \label{fig:PlotDotRadDensVaryk}
\end{figure}

At the strong external potential $k=8$, corresponding to steep gradients,
\figref{fig:QntmDotVaryK} shows the DFT energy is overestimated compared with
the DMC result, indicating that the gradient approximation is not applicable and
that the next order term in a gradient expansion is
negative. \figref{fig:PlotDotRadDensVaryk} shows that the DFT density profile
underestimates the true density towards the center of the dot and overestimates
density in the outer regions, indicating that the DFT functional does not favor
steep enough gradients. This is consistent with the next term in the gradient expansion
being negative. The breakdown corresponds to a coefficient of $\alpha\approx1.8$
in $q_{\text{max}}=\alpha(\hbar a_{0}^{*-1/4})n^{1/4}$, close to the
$\alpha\approx4$ which corresponds to the maximum contribution to the
interacting energy.

At the intermediate potential $k=1$ the DFT and DMC estimates of energy and
the density profile agree, in this region the gradient approximation
applies. The DFT density profile shows a slight over-density at the center,
consistent with self-interaction energy being included in the DFT
calculation. At the weak potential $k=0.1$ electron densities are low meaning
the homogeneous interacting energy is outside of its region of applicability
(see \figref{fig:CASINOhomogeneousgas}), therefore the DFT energy is an
overestimate.

\subsubsection{Varying the gradient term coefficient $\xi$}

\begin{figure}
 \includegraphics{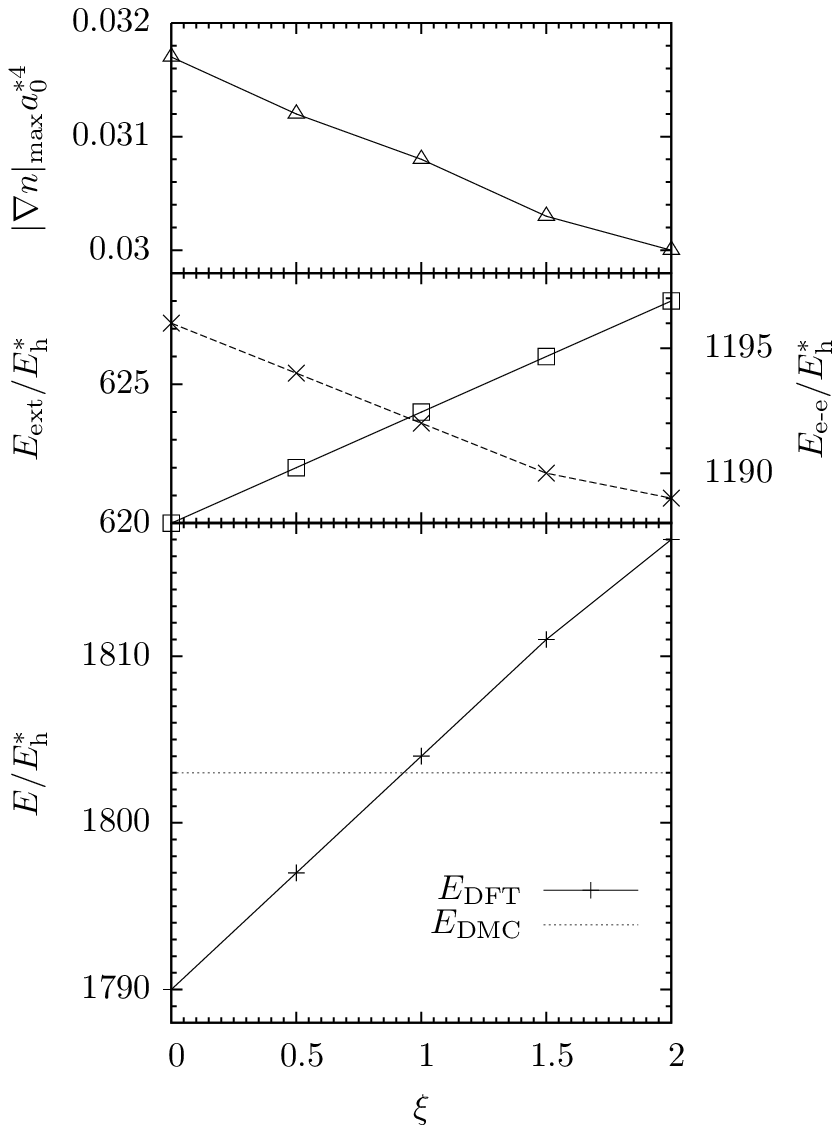}
 \caption{The upper panel shows, for dots with external potential strength
 $k=1$, the density profile maximum gradient as a function of
 $\xi$. The central panel shows the variation of external
 potential DFT energy based on the primary $y$-axis using square points and the solid line,
 the secondary $y$-axis shows electron-electron DFT energy using crosses and the
 dashed line. The lower panel solid line shows the variation of DFT ground state
 energy with $\xi$, and the horizontal dotted line the ground-state energy
 predicted using DMC from the $\xi=1$ trial wave function. The DMC statistical
 error is less than the size of the points.}
 \label{fig:QntmDotVaryChi}
\end{figure}

\figref{fig:QntmDotVaryChi} shows results of simulations on dots, chosen to have
a potential strength $k=1$, which is at the center of agreement of the previous
results. The best agreement between the DFT and DMC ground state energy is at
$\xi\sim0.9$. This is in good agreement with the expected $\xi=1$, the
difference may be due to systematic errors such as the self-interacting energy
or higher order gradient terms. As expected, the energy is overestimated for
dots with too large a gradient expansion term, and underestimated for dots with
too small a gradient correction term.

\begin{figure}
 \includegraphics{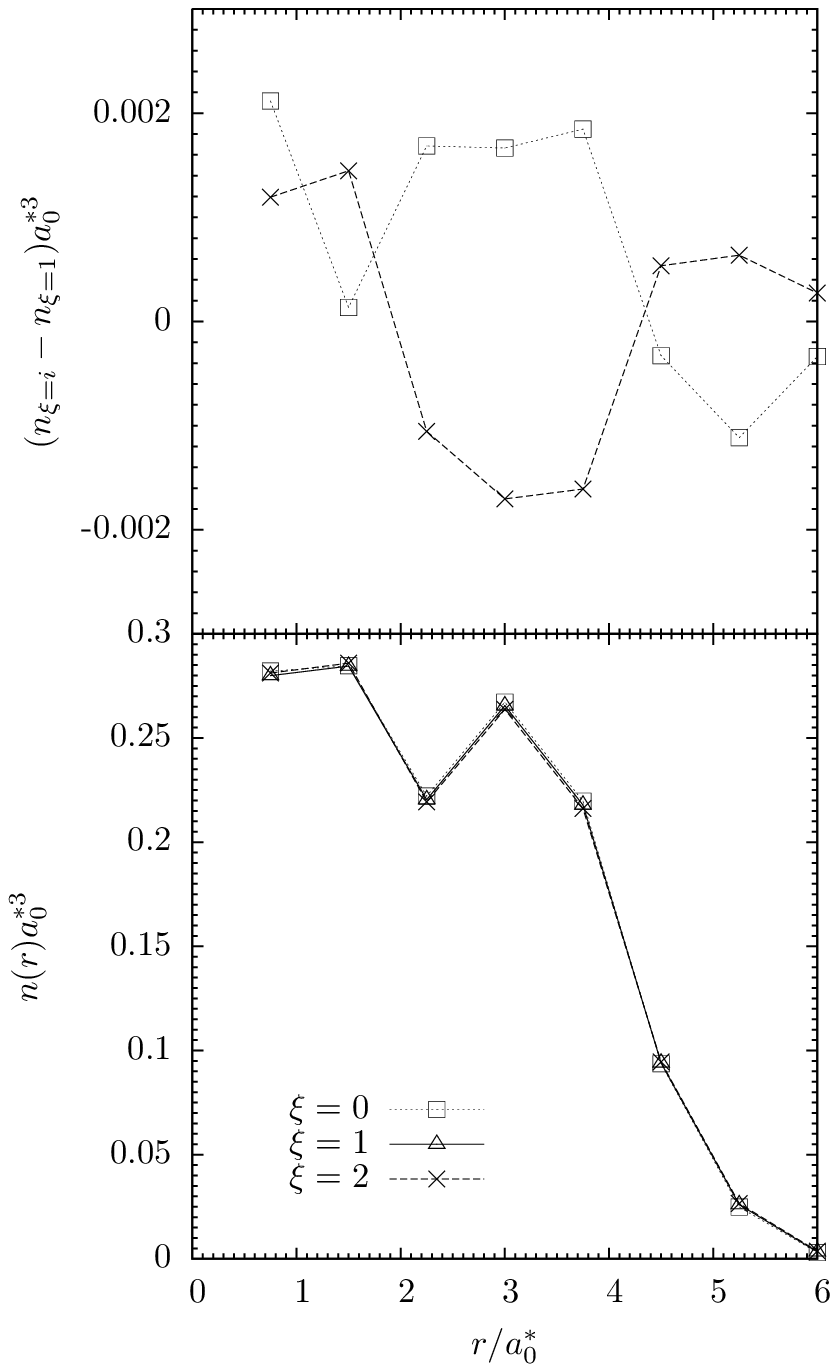}
 \caption{The upper panel shows the variation of the $\xi=0$ (dotted line,
 squares) and the $\xi=2$ (dashed line, crosses) DFT density profiles from the
 $\xi=1$ DFT density profile. The lower panel shows DFT density profiles for
 $\xi=0$, $\xi=1$ (solid line, triangles), and $\xi=2$ in a $k=1$ dot.}
 \label{fig:PlotDotRadDensVaryChi}
\end{figure}

The maximum gradient seen in the dot density profile decreases as $\xi$
increases (see \figref{fig:QntmDotVaryChi}). The dot becomes more spread out so
the external energy $E_{\text{ext}}$ increases whilst the total
electron-electron Coulomb energy $E_{\text{e-e}}$ decreases. Overall the total
DFT energy increases. Three quantum dot electron density profiles for gradient
term coefficients $\xi=1$, 2, and 3 are shown in
\figref{fig:PlotDotRadDensVaryChi}. Compared with the dot calculated with
$\xi=1$, the dot generated with no energy penalty for gradients, $\xi=0$, has a
high central and low outer density showing that it has a higher gradient in the
density. Conversely dot with increased energy cost for gradients, $\xi=2$, has a
more shallow profile.

The density profiles seen in \figref{fig:PlotDotRadDensVaryk} and
\figref{fig:PlotDotRadDensVaryChi} can be further analyzed in light of other
theoretical studies of quantum dots reviewed in Ref.~\cite{02rm10}. The density
profile calculated using the many-flavor functional is not flat at the center,
but instead has correlation-induced density inhomogeneity evidenced by a
characteristic minimum in the density at $r\approx2a_{0}^{*}$. The
intermediate density regime in which this occurs is consistent with the strong
correlations causing a minimum in the total many-flavor energy density
\cite{08c07}. It is also akin to the intermediate density regime seen in other
quantum dot systems \cite{99ehmg04,02rm10,06gguub05,08krpgpgmpw06}, in the high
density limit the quantum dot has properties like a Fermi liquid with
de-localized electrons \cite{99ehmg04,02rm10,05gv04}, whereas in the low density
limit the electrons become crystalline
\cite{90lm04,92lm06,93br03,98mvn07,99ehmg04,02rm10} inside the dot. As the
many-flavor functional was successful in predicting correlation-induced
inhomogeneities, it could be used to investigate other many-flavor quantum dot
effects including the Kondo effect in multi-valley semiconductors
\cite{07scj05,07sj11}, the reduction of valley degeneracy of coupled quantum
dots \cite{03he10,05a08,07bb07}, and harmonically trapped cold atoms with an
additional quantum number denoting energy level
\cite{04hh04,04hh09,07crd09,07yl11}.

\section{Conclusions}

We have computationally verified the theory of the MFEG presented in
\cite{08c07} using QMC simulations. In a homogeneous system, DMC estimates for
the ground state energy are consistent with theory and the theoretically
estimated density range over which the theory applies is consistent with
numerical results. The applicable density for $\text{Bi}_{2}\text{Te}_{3}$
$(\nu=6)$ corresponds to a charge carrier density between
$7\times10^{19}\unit{cm}^{-3}$ and $2\times10^{20}\unit{cm}^{-3}$.

The density response function for a MFEG with 24 flavors was found using three
methods: density modulation predicted by VMC, and the variation in ground state
energy predicted by VMC and also by DMC. The two VMC results underestimated the
response $1/\epsilon$, but the DMC results agreed with theory and could
distinguish between the exact and a useful approximate expression for
polarizability.

We used a many-flavor functional including a local gradient approximation in DFT
calculations of large quantum dots. The DFT calculation estimated the
ground-state energy and wave function, which were verified by a DMC
calculation. We found the high gradient breakdown of the expansion was at
$q_{\text{max}}\approx1.8(\hbar a_{0}^{*-1/4})n^{1/4}$, the low gradient
breakdown was consistent with the homogeneous MFEG lowest applicable density,
and that the gradient expansion was applicable in the intermediate regime. The
many-flavor functional, used as part of DFT calculations, could be a useful tool
for analyzing other multi-valley semiconductor systems.

\begin{acknowledgements}
G.J.C. acknowledges the financial support of an EPSRC studentship, P.D.H. was
supported by a Royal Society University Research Fellowship. We thank
N.D.M.~Hine for providing the DOTDFT code, P.~L\'opez R\'ios and N.D.~Drummond
for help modifying and running CASINO, R.~Needs for providing computing time, and
A.J.~Morris for careful reading of the manuscript.
\end{acknowledgements}


\end{document}